\RequirePackage{fixltx2e}
\documentclass[aip,apl,reprint]{revtex4-1}
\usepackage{graphicx}
\usepackage{amsmath}

\draft % marks overfull lines with a black rule on the right

\begin{document}

%\preprint{}
%COPS INTERNAL REPORT SS141010      
\title{Local thermal resonance control of GaInP photonic crystal membrane cavities using ambient gas cooling} %Title of paper


\author{Sergei Sokolov}
\email[]{s.sokolov@utwente.nl}
\homepage[]{http://cops.nano-cops.com}
\affiliation{Complex Photonic Systems (COPS), MESA+ Institute for Nanotechnology, University of Twente,7500 AE Enschede, The Netherlands}
\author{Jin Lian}
\affiliation{Complex Photonic Systems (COPS), MESA+ Institute for Nanotechnology, University of Twente,7500 AE Enschede, The Netherlands}
\author{Emre Y\"uce}
\affiliation{Complex Photonic Systems (COPS), MESA+ Institute for Nanotechnology, University of Twente,7500 AE Enschede, The Netherlands}
\author{Sylvain Combri\'{e}}
\affiliation{Thales Research \& Technology, Route D\'{e}partementale 128, 91767 Palaiseau, France}
\author{Gaelle Lehoucq}
\affiliation{Thales Research \& Technology, Route D\'{e}partementale 128, 91767 Palaiseau, France}
\author{Alfredo De Rossi}
\affiliation{Thales Research \& Technology, Route D\'{e}partementale 128, 91767 Palaiseau, France}
\author{Allard P. Mosk}
\affiliation{Complex Photonic Systems (COPS), MESA+ Institute for Nanotechnology, University of Twente,7500 AE Enschede, The Netherlands}
%\thanks{}
\makeatletter
\let\setrefcountdefault\relax
\makeatother
%\date{SS141010 \today}
\date{\today}

\begin{abstract}
We perform spatially dependent tuning of a GaInP photonic crystal cavity using a continuous wave violet laser. Local tuning is obtained by laser heating of the photonic crystal membrane. The cavity resonance shift is measured for different pump positions and for two ambient gases: He and N\textsubscript{2}. We find that the width of the temperature profile induced in the membrane depends strongly on the thermal conductivity of the ambient gas. For He gas a narrow spatial width of the temperature profile of 2.8 $\mu$m is predicted and verified in experiment. 
%Such localized heating is useful for thermal control of high-density photonic integrated circuits and coupled resonator optical waveguides.
\end{abstract}


\maketitle %\maketitle must follow title, authors, abstract and \pacs

%\section{}
 %\label{}
%\subsection{}
%\subsubsection{}
Photonic crystal (PhC) cavities are widely studied because of their fascinating applications\cite{yariv, noda, matsko}. Arrays of PhC cavities can form coupled resonator optical waveguides (CROW), which are very promising for slow light applications\cite{notomi200} and the study of light localization\cite{localiz}. Various fabrication imperfections can cause a disorder in a CROW structure which leads to a detuning of cavities from the intended resonance and reduces the waveguide throughput and bandwidth. Tuning each cavity independently can restore cavities in resonance and counteract the disorder. There is a variety of methods to change the refractive index of a PhC cavity, including  free-carrier injection\cite{freecarrier}, the nonlinear Kerr effect\cite{kerr}, thermal effects \cite{notomitherm,stanfordtherm}, oxidation\cite{oxydation1,oxydation2,oxydation3} and chemical processes\cite{chem}. Of these methods, thermal tuning through local laser heating is easy, reversible and can give a steady-state control of the resonance properties of the system. However due to the diffusion of heat in the PhC membrane thermal control of one cavity will affect the neighbor cavities. The width of the temperature profile is determined by the sample material and the surrounding media. Therefore one can expect that by carefully selecting the  sample material and ambient medium one can control the width of the temperature profile. 

In this work we use the semiconductor alloy Ga$_{0.51}$In$_{0.49}$P as a sample material and two gases, nitrogen and helium, as a surrounding media. Ga$_{0.51}$In$_{0.49}$P has a thermal conductivity\cite{adachi} of 4.9 W/(m$\cdot$K) which is quite small in comparison to other semiconductor materials\cite{inp}.  The thermal conductivity of gases is often assumed to be negligible compared to semiconductors, such as Si. 
However, the effect of the gas on the width of the thermal profile depends strongly on the ratio of the thermal conductivity of the gas and the semiconductor.
Helium has a high thermal conductivity\cite{waterhydrogen} of 0.153 W/(m$\cdot$K), which is more than 6 times higher than of nitrogen\cite{waterhydrogen}(0.024 W/(m$\cdot$K)) and only 32 times smaller that that of Ga$_{0.51}$In$_{0.49}$P. Therefore the combination of Ga$_{0.51}$In$_{0.49}$P and He should have a high thermal exchange efficiency and small  width of the temperature profile in comparison with other materials. To investigate the width of the temperature profile in PhC membranes we measured the response of the  resonance of a H0-type cavity to a spatially scanned continuous wave (CW) heating laser focused on the membrane. 
\begin{figure}[h]
 \includegraphics[width=8cm]{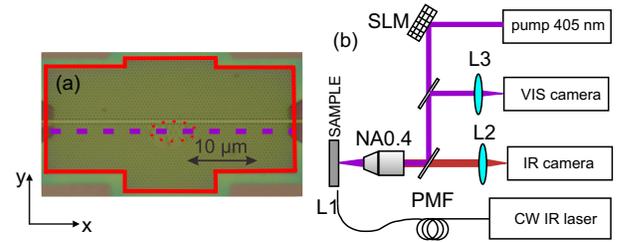}
 \caption{\label{FIG1}(a) Optical microscope image of a GaInP PhC (red line) with an H0 cavity (red dotted ellipse) in the center. (b) The pump-probe setup. }%
 \end{figure}
 
The PhC H0 nanocavity is implemented in a triangular lattice of air holes with a period $a$=505 nm and a hole radius 0.24 $a$. The hole shift for the H0 cavity is 0.16 $a$. A subharmonic hole structure is induced to enhance the radiation from the cavity in a vertical direction\cite{bandfolding}. The cavity is made of a suspended membrane of  Ga$_{0.51}$In$_{0.49}$P with a thickness of 180 nm. The membrane is covered with 30 nm of Si$_{3}$N$_{4}$ on the top using a PECVD technique to tune the cavity resonance. The detailed description of the fabrication method can be found in Ref.\citenum{fabrication}. A microscope picture of the cavity with the surrounding PhC is presented in Fig.~\ref{FIG1}a. The cavity is close to the center of the membrane and the IR light is evanescently coupled through a PhC waveguide. Mode converters at the end of the waveguide decrease the Fabry-Perot interference in the waveguide and increase the coupling efficiency\cite{modeconverter}.

 The setup is shown in Fig.~\ref{FIG1}b. The CW IR laser probe light was coupled to the PhC waveguide using a polarization maintaining lensed fiber with NA of 0.55. The out-of-plane scattered light was collected using a 0.4 NA objective and imaged on an IR CCD camera using tube lens L2. The total magnification of the system is 50x. We recorded spectra by sweeping the wavelength of the IR laser and taking IR camera snapshots for each wavelength. The oxygen concentration was reduced to  less than 0.025\% to suppress oxidation effects\cite{oxydation1,oxydation2,oxydation3}. The chamber was kept at a slight overpressure (less than 2 mbar above atmospheric pressure). Thermal tuning was performed using a 405 nm CW diode laser, which was focused onto the surface of the sample into a spot with a FWHM of 0.96 $\mu$m using the same objective. The surface of the sample with the pump spot was also imaged with a visible range camera using tube lens L3 with a system magnification of 27x. The position of the pump spot was controlled using a spatial light modulator. The pump spot was moved along a line through the center of the cavity (see Fig.~\ref{FIG1}a) and the resonance spectrum of the cavity was measured for a sequence of pump positions. Any effects of slow oxidation or water coverage are eliminated by alternating between reference spectra (without pump laser) and signal spectra (with pump).
  \begin{figure} [h]
 \includegraphics[width=7.8cm]{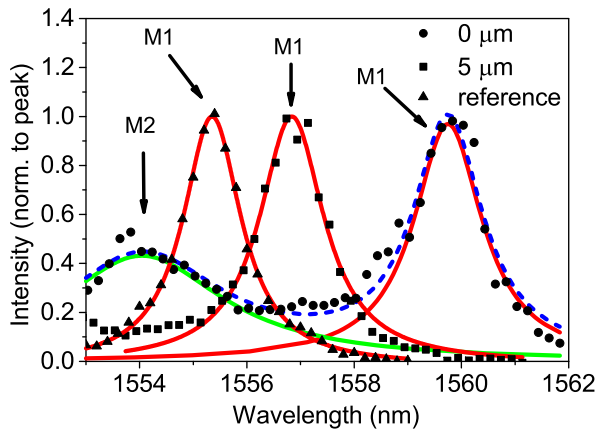}
\caption{\label{FIG2} Spectra obtained from the out-of-plane scattered light for no-pump reference and for pump positions on top of the cavity and 5 $\mu$m away from the center of the cavity in $\mathrm{N}_{2}$ atmosphere. Color lines are Lorentzian fits.}%
  \end{figure}
  
Resonance spectra for the reference and for pump positions of 0 and 5 $\mu$m relative to the center of the cavity in N$_2$ are presented in Fig.~\ref{FIG2}.  The closer the pump spot is placed to the cavity, the stronger the redshift is due to the better overlap of the temperature profile and the cavity mode. We observe two resonance modes: the high-\textit{Q} low-frequency mode (M1) and the low-\textit{Q} high-frequency M2.  Both resonances have a distinct Lorentzian shape and quality factors of about 1000 and 300 are estimated for modes M1 and M2 respectively. The calculated loaded \textit{Q}-factor of the mode M1 is about 2500, because the cavity is designed to be overcoupled. The resonance M2 corresponds to a higher order mode which is predicted from calculations. The mode M1 is much brighter than M2. The two modes have slightly different spatial positions, and the reference position is the center of mode M1. We use the wavelength shift of mode M1 to determine the spatially dependent cavity response to the pump. Apart from the redshift we observed a reduction of the intensity of the out-of-plane scattered light when the pump was positioned close to the cavity. We attribute this reduction to the fact that the cavity resonance was tuned farther away from the PhC band, thereby reducing the evanescent coupling to the waveguide.  

\begin{figure} [h]
 \includegraphics[width=8cm]{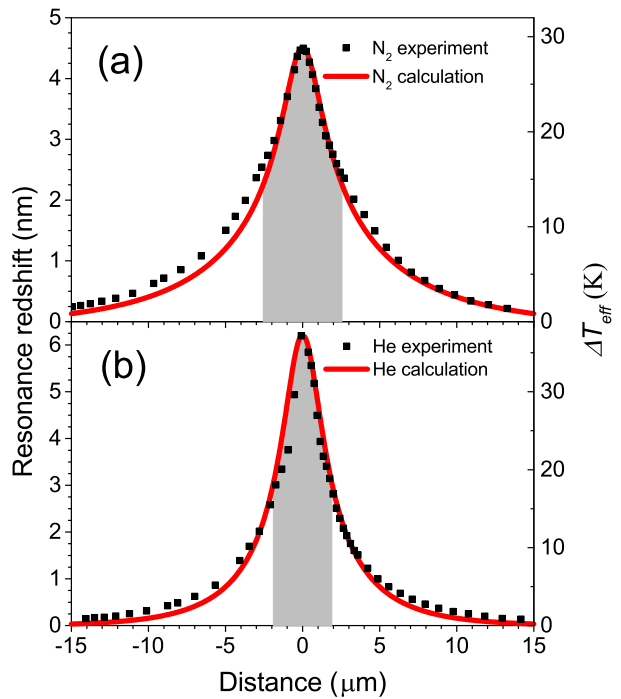}
\caption{\label{FIG3}Redshift of the cavity resonance and $\Delta T_{eff}$ for different pump positions with nitrogen (a) and helium (b) as an ambient gas. Red curves represent the redshift calculated according to the model. From one measurement to another we observed an error less than 3\% of the shift for $\mathrm{N}_{2}$, and less than 1\% for He.}
  \end{figure}

The pump power incident on the sample was 110 $\mu$W in $\mathrm{N}_{2}$ atmosphere. The spatial response curve for $\mathrm{N}_{2}$ atmosphere is presented in Fig.~\ref{FIG3}a.        
The maximum redshift is 4.5 nm and the FWHM spatial width ($\Delta x_{eff}$) of the tuning curve is 5.1 $\mu$m.
 In He atmosphere a larger pump power of 220 $\mu$W was used (see Fig.~\ref{FIG3}b). The maximum observed redshift is 6.2 nm and $\Delta x_{eff}$ is only 3.7 $\mu$m.  As a result of the high thermal conductivity of the He, both the redshift (normalized to power) and $\Delta x_{eff}$ are smaller by 30\% compared to $\mathrm{N}_{2}$. 
This thermal profile can be used to predict thermal crosstalk in structures with multiple cavities such as CROWs. Compensation of crosstalk is typically possible when the crosstalk is below about 50\%, i.e. for elements spaced more than $\Delta x_{eff}/2$. A reduction of $\Delta x_{eff}$ therefore directly translates into a higher possible integration density of independently tuned cavities.
 
We have performed numerical calculations to model the resonance wavelength change of the cavity for different pump positions taking into account material and pump properties. For small perturbations of the refractive index the wavelength change can be expressed as\cite{joannopoulos}
  \begin{align}\label{eq:wlchange}
  \Delta\lambda&=\frac{\lambda}{\langle n\rangle}\frac{\mathrm dn}{\mathrm dT}\Delta T_{eff}, \mathrm{with}\\
 \Delta T_{eff}&=\frac{\int\Delta T(\mathbf{r}-\mathbf{r_{0}})\epsilon (\mathbf r)|\mathbf{E}(\mathbf r)|^2d^{3} r}{\int\epsilon (\mathbf r)|\mathbf{E}(\mathbf r)|^2d^{3}r},
 \end{align}
where$\Delta T_{eff}$ is temperature averaged over the mode profile $\epsilon (\mathbf r)|\mathbf{E}(\mathbf r)|^2$ of the cavity, $\epsilon(\mathbf r)$ is the dielectric constant of the membrane material,  $\Delta T=T-T_0$ is the local temperature change in the PhC averaged over the membrane thickness,  $\frac{\mathbf dn}{\mathbf dT}$ is the linear coefficient of the refractive index temperature response, $\langle n\rangle$ is the averaged refractive index of the membrane material and $\mathbf{r_{0}}$ is the position of the pump. $T_{0}=293.15 K$ is the ambient temperature.  The mode profile was calculated using a finite difference time domain method\cite{meep}. Temperature profiles were calculated in COMSOL using a steady state heat diffusion model with a constant source term assuming that thermal conductivities are not temperature dependent within the relevant temperature range: 
  \begin{equation}
 \kappa( x,y,z ){\nabla}^2 T( x,y,z ) + W(x,y,z)=0,
 \end{equation}
  where $\kappa( x,y,z )$ is the thermal conductivity and $W$ is a source term inside the membrane written as a sum of two terms $W=W_{E}+W_{R}$, due to the excess energy of the carriers and their recombination respectively.  
 
  We observe photoluminescence (PL) of  Ga$_{0.51}$In$_{0.49}$P at 670 nm caused by the pump light. The bandgap of Ga$_{0.51}$In$_{0.49}$P is 1.85 eV and the energy of a  405 nm photon is 3.06 eV. The excess energy term ($W_{E}$) is released within about a picosecond upon optical excitation and therefore has the spatial profile of the pump spot (a Gaussian width $\sigma_E=0.41 \, \mu$m) and corresponds to a fraction $f_E=0.4$ of the initial power. The recombination term ($W_{R}$) has the profile of the PL spot, a Gaussian width $\sigma_R=0.50 \,\mu$m, and it takes a fraction $f_R=0.6$ of the photon energy. We assume that all absorbed light is converted into heat, because the estimated radiative carrier recombination efficiency is less than 1\%. 
The excess energy term is expressed as
\begin{equation}
W_E(x,y,z)=f_E\frac{P(1-R)}{2\pi \sigma_E^{2}}\alpha \mathrm e^{-(x^2+y^2)/2\sigma_E^{2}}\mathrm e^{-\alpha z}.
\end{equation}
 Here R is the reflection coefficient of the membrane estimated to be 9.5$\%$, $\alpha$ is the absorption coefficient of Ga$_{0.51}$In$_{0.49}$P, and $P$ is the total power of the pump beam. The recombination heating term has the same Gaussian shape with the appropriate fraction and width inserted.
 The size of the air gap between the membrane and GaAs substrate for our samples is equal to 1.5 $\mu$m, the size of the air layer above the membrane was taken to be 10 $\mu$m. All external boundaries are Dirichlet type with temperature $T_0$. 
The photonic crystal is assumed to be a rectangle of 40$\times$20 $\mu$m. Dirichlet  boundaries are considered because the membrane is surrounded from four sides by the bulk material, which acts like a perfect heat sink. Thermal conductivities for gases were taken from Ref.\citenum{waterhydrogen}. The coating layer made of Si$_{3}$N$_{4}$ has a thermal conductivity\footnote{MIT material property database, http://www.mit.edu/~6.777/matprops/matprops.htm} of 24.5 W/(m$\cdot$K). The thermal conductivity of the membrane was taken as volume averaged between holes of the PhC lattice and two layers of Si$_{3}$N$_{4}$ and Ga$_{0.51}$In$_{0.49}$P. The absorption coefficient \cite{schubert} of Ga$_{0.51}$In$_{0.49}$P at 405 nm is 24.5 $(\mathrm{\mu m})^{-1}$ and the absorption coefficient\cite{sin} of Si$_{3}$N$_{4}$ is less than 0.006  $(\mathrm{\mu m})^{-1}$. As a result nearly all pump light is absorbed by the Ga$_{0.51}$In$_{0.49}$P membrane. The temperature profile can be calculated without any free parameters. The only parameter which is unknown for the redshift calculation is $\frac{dn}{dT}$ for Ga$_{0.51}$In$_{0.49}$P.

The resulting effective temperature profiles are presented in Fig.~\ref{FIG3}a,b, and are in excellent agreement with the experiment for both nitrogen and helium. The maximum temperature rise in case of nitrogen atmosphere is 29 K, and in case of He it is 37 K, so the ratio of these two values normalized to the incident power is equal to 0.63. The experimental ratio of redshifts is 0.69.  From Fig.~\ref{FIG3} we also estimate the so far unknown $\frac{dn}{dT}$ for Ga$_{0.51}$In$_{0.49}$P. We obtain the value of $3.1\pm0.5\times10^{-4} \mathrm{K^{-1}}$. 
\begin{figure}[h]
  \includegraphics[width=8cm]{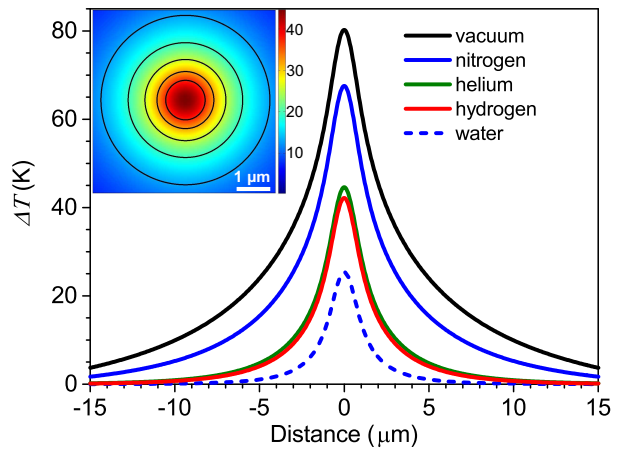}
  \caption{\label{FIG4}
X-axis crossection of the calculated temperature profile ($\Delta T$) for different ambient media. The pump focus is placed in the center of the cavity. The power of the pump light is  220 $\mu$W. The inset shows the 2D averaged temperature profile for helium. The isotherms have a step of \mbox{6.3 K}.}
  \end{figure}

Instead of helium one could use media with a higher thermal conductivity. In Fig.~\ref{FIG4} calculated temperature profiles  along the central line  of the PhC membrane (see Fig.~\ref{FIG1}) are presented for different surrounding media. These are bare temperature profiles, not convolved with the cavity mode. 
The widest temperature profile is found in case of vacuum. Water has a thermal conductivity 4 times higher than He, which results in a significantly narrower temperature profile. We note that low-refractive-index liquids such as acetone and isopropanol have thermal conductivities comparable to that of He. Using liquids for heat exchange completely changes the optical properties of the sample, therefore using a high-thermal-conductivity gas is a more appropriate option in many cases. 
  
\begin{figure}[h]
  \includegraphics[width=8cm]{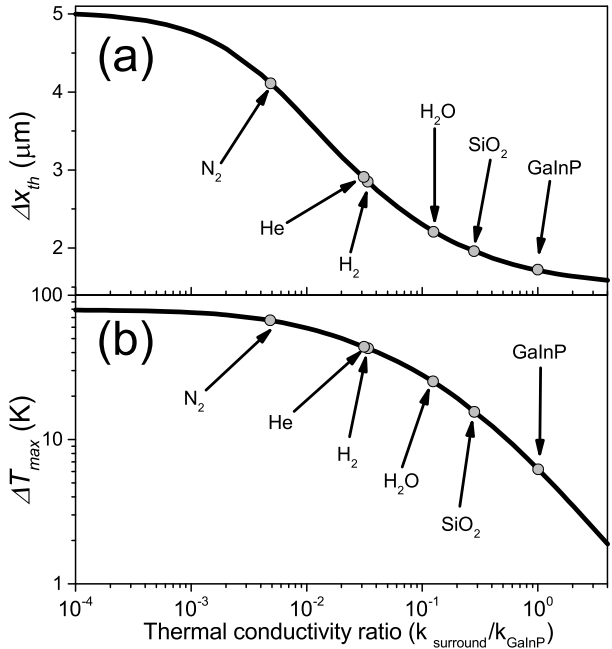}
  \caption{\label{FIG5} Calculated $\Delta x_{th}$ (a) and $\Delta T_{max}$  (b) vs thermal conductivity ratio of surrounding media and Ga$_{0.51}$In$_{0.49}$P. Some well-known materials are marked. The power of the pump light is 220 $\mu$W.}
  \end{figure}

The general tendency of the width $\Delta x_{th}$ and the peak  value ($\Delta T_{max}$) of the bare temperature profile is presented in Fig.~\ref{FIG5}.
 In case of low ambient thermal conductivity heat tends to diffuse inside the membrane, therefore $\Delta x_{th}$ and the peak temperature reach constant values of 5 $\mu$m and 79 K, respectively. For large thermal conductivities of the surrounding media all heat diffuses rapidly into the ambient media reducing $\Delta x_{th}$ and the peak temperature.   
  
  Gas cooling is most relevant for thermal tuning of resonances in membranes made of materials with small thermal conductivities such as Ga$_{0.51}$In$_{0.49}$P. In highly conducting materials like Si the thermal profile is significantly larger\cite{stanfordtherm} and is typically determined by the material boundaries rather than by heat exchange with the gas. 
      
  For time-dependent tuning applications such as a switch or a tracking filter the time response is of importance. From time-dependent simulations we estimate that the 10\% to 90\% temperature rise time in a nitrogen atmosphere is about 6 $\mu$s, but in  helium it reduces to  2 $\mu$s. Therefore, helium not only helps to improve the spatial resolution of thermal tuning but also one makes the tuning response about three times faster. Similarly, it should reduce thermal optical nonlinearities that result from absorption of cavity-enhanced probe light.
  
  In conclusion, we measured the spectral tuning curve of a thermally tuned cavity for different surrounding media and found an excellent agreement between experiment and our numerical calculations. Our results and our calculations are of importance for applications such as thermal control of CROWs where the width of the temperature profile should be minimized. Also our  calculations offer insight in the role of  ambient materials for any applications that involve local laser heating.  
  
  The authors thank Henri Thyrrestrup and Willem Vos for helpful discussions and advice and Cornelis Harteveld for technical support. This work was supported by the European Research Council project No. 279248.


\begin{thebibliography}{24}%
\makeatletter
\providecommand \@ifxundefined [1]{%
 \@ifx{#1\undefined}
}%
\providecommand \@ifnum [1]{%
 \ifnum #1\expandafter \@firstoftwo
 \else \expandafter \@secondoftwo
 \fi
}%
\providecommand \@ifx [1]{%
 \ifx #1\expandafter \@firstoftwo
 \else \expandafter \@secondoftwo
 \fi
}%
\providecommand \natexlab [1]{#1}%
\providecommand \enquote  [1]{``#1''}%
\providecommand \bibnamefont  [1]{#1}%
\providecommand \bibfnamefont [1]{#1}%
\providecommand \citenamefont [1]{#1}%
\providecommand \href@noop [0]{\@secondoftwo}%
\providecommand \href [0]{\begingroup \@sanitize@url \@href}%
\providecommand \@href[1]{\@@startlink{#1}\@@href}%
\providecommand \@@href[1]{\endgroup#1\@@endlink}%
\providecommand \@sanitize@url [0]{\catcode `\\12\catcode `\$12\catcode
  `\&12\catcode `\#12\catcode `\^12\catcode `\_12\catcode `\%12\relax}%
\providecommand \@@startlink[1]{}%
\providecommand \@@endlink[0]{}%
\providecommand \url  [0]{\begingroup\@sanitize@url \@url }%
\providecommand \@url [1]{\endgroup\@href {#1}{\urlprefix }}%
\providecommand \urlprefix  [0]{URL }%
\providecommand \Eprint [0]{\href }%
\providecommand \doibase [0]{http://dx.doi.org/}%
\providecommand \selectlanguage [0]{\@gobble}%
\providecommand \bibinfo  [0]{\@secondoftwo}%
\providecommand \bibfield  [0]{\@secondoftwo}%
\providecommand \translation [1]{[#1]}%
\providecommand \BibitemOpen [0]{}%
\providecommand \bibitemStop [0]{}%
\providecommand \bibitemNoStop [0]{.\EOS\space}%
\providecommand \EOS [0]{\spacefactor3000\relax}%
\providecommand \BibitemShut  [1]{\csname bibitem#1\endcsname}%
\let\auto@bib@innerbib\@empty
%</preamble>
\bibitem [{\citenamefont {Yariv}\ \emph {et~al.}(1999)\citenamefont {Yariv},
  \citenamefont {Xu}, \citenamefont {Lee},\ and\ \citenamefont
  {Scherer}}]{yariv}%
  \BibitemOpen
  \bibfield  {author} {\bibinfo {author} {\bibfnamefont {A.}~\bibnamefont
  {Yariv}}, \bibinfo {author} {\bibfnamefont {Y.}~\bibnamefont {Xu}}, \bibinfo
  {author} {\bibfnamefont {R.~K.}\ \bibnamefont {Lee}}, \ and\ \bibinfo
  {author} {\bibfnamefont {A.}~\bibnamefont {Scherer}},\ }\href@noop {}
  {\bibfield  {journal} {\bibinfo  {journal} {Opt. Express}\ }\textbf {\bibinfo
  {volume} {24}},\ \bibinfo {pages} {711} (\bibinfo {year} {1999})}\BibitemShut
  {NoStop}%
\bibitem [{\citenamefont {Akahane}\ \emph {et~al.}(2003)\citenamefont
  {Akahane}, \citenamefont {Asano}, \citenamefont {Song},\ and\ \citenamefont
  {Noda}}]{noda}%
  \BibitemOpen
  \bibfield  {author} {\bibinfo {author} {\bibfnamefont {Y.}~\bibnamefont
  {Akahane}}, \bibinfo {author} {\bibfnamefont {T.}~\bibnamefont {Asano}},
  \bibinfo {author} {\bibfnamefont {B.}~\bibnamefont {Song}}, \ and\ \bibinfo
  {author} {\bibfnamefont {S.}~\bibnamefont {Noda}},\ }\href@noop {} {\bibfield
   {journal} {\bibinfo  {journal} {Nature}\ }\textbf {\bibinfo {volume}
  {425}},\ \bibinfo {pages} {944} (\bibinfo {year} {2003})}\BibitemShut
  {NoStop}%
\bibitem [{\citenamefont {Matsko}(2009)}]{matsko}%
  \BibitemOpen
  \bibfield  {author} {\bibinfo {author} {\bibfnamefont {A.~B.}\ \bibnamefont
  {Matsko}},\ }\href@noop {} {\emph {\bibinfo {title} {Practical Applications
  of microresonators in optics and photonics}}}\ (\bibinfo  {publisher} {CRC
  Press},\ \bibinfo {year} {2009})\BibitemShut {NoStop}%
\bibitem [{\citenamefont {Matsuda}\ \emph {et~al.}(2014)\citenamefont
  {Matsuda}, \citenamefont {Kuramochi}, \citenamefont {Takesue},\ and\
  \citenamefont {Notomi}}]{notomi200}%
  \BibitemOpen
  \bibfield  {author} {\bibinfo {author} {\bibfnamefont {N.}~\bibnamefont
  {Matsuda}}, \bibinfo {author} {\bibfnamefont {E.}~\bibnamefont {Kuramochi}},
  \bibinfo {author} {\bibfnamefont {H.}~\bibnamefont {Takesue}}, \ and\
  \bibinfo {author} {\bibfnamefont {M.}~\bibnamefont {Notomi}},\ }\href@noop {}
  {\bibfield  {journal} {\bibinfo  {journal} {Opt. Lett.}\ }\textbf {\bibinfo
  {volume} {39}},\ \bibinfo {pages} {2290} (\bibinfo {year}
  {2014})}\BibitemShut {NoStop}%
\bibitem [{\citenamefont {Mookherjea}\ \emph {et~al.}(2008)\citenamefont
  {Mookherjea}, \citenamefont {Park}, \citenamefont {Yang},\ and\ \citenamefont
  {Bandaru}}]{localiz}%
  \BibitemOpen
  \bibfield  {author} {\bibinfo {author} {\bibfnamefont {S.}~\bibnamefont
  {Mookherjea}}, \bibinfo {author} {\bibfnamefont {J.~S.}\ \bibnamefont
  {Park}}, \bibinfo {author} {\bibfnamefont {S.}~\bibnamefont {Yang}}, \ and\
  \bibinfo {author} {\bibfnamefont {P.~R.}\ \bibnamefont {Bandaru}},\
  }\href@noop {} {\bibfield  {journal} {\bibinfo  {journal} {Nat. Photonics}\
  }\textbf {\bibinfo {volume} {2}},\ \bibinfo {pages} {90} (\bibinfo {year}
  {2008})}\BibitemShut {NoStop}%
\bibitem [{\citenamefont {Husko}\ \emph {et~al.}(2009)\citenamefont {Husko},
  \citenamefont {\mbox{De Rossi}}, \citenamefont {Combri\'{e}}, \citenamefont
  {Tran}, \citenamefont {Raineri},\ and\ \citenamefont {Wong}}]{freecarrier}%
  \BibitemOpen
  \bibfield  {author} {\bibinfo {author} {\bibfnamefont {C.}~\bibnamefont
  {Husko}}, \bibinfo {author} {\bibfnamefont {A.}~\bibnamefont {\mbox{De
  Rossi}}}, \bibinfo {author} {\bibfnamefont {S.}~\bibnamefont {Combri\'{e}}},
  \bibinfo {author} {\bibfnamefont {Q.~V.}\ \bibnamefont {Tran}}, \bibinfo
  {author} {\bibfnamefont {F.}~\bibnamefont {Raineri}}, \ and\ \bibinfo
  {author} {\bibfnamefont {C.~W.}\ \bibnamefont {Wong}},\ }\href@noop {}
  {\bibfield  {journal} {\bibinfo  {journal} {Appl. Phys. Lett.}\ }\textbf
  {\bibinfo {volume} {94}},\ \bibinfo {pages} {021111} (\bibinfo {year}
  {2009})}\BibitemShut {NoStop}%
\bibitem [{\citenamefont {Y$\mathrm{\ddot{u}}$ce}\ \emph
  {et~al.}(2013)\citenamefont {Y$\mathrm{\ddot{u}}$ce}, \citenamefont
  {Ctistis}, \citenamefont {Claudon}, \citenamefont {Dupuy}, \citenamefont
  {Buijs}, \citenamefont {\mbox{de Ronde}}, \citenamefont {Mosk}, \citenamefont
  {G\'{e}rard},\ and\ \citenamefont {Vos}}]{kerr}%
  \BibitemOpen
  \bibfield  {author} {\bibinfo {author} {\bibfnamefont {E.}~\bibnamefont
  {Y$\mathrm{\ddot{u}}$ce}}, \bibinfo {author} {\bibfnamefont {G.}~\bibnamefont
  {Ctistis}}, \bibinfo {author} {\bibfnamefont {J.}~\bibnamefont {Claudon}},
  \bibinfo {author} {\bibfnamefont {E.}~\bibnamefont {Dupuy}}, \bibinfo
  {author} {\bibfnamefont {R.~D.}\ \bibnamefont {Buijs}}, \bibinfo {author}
  {\bibfnamefont {B.}~\bibnamefont {\mbox{de Ronde}}}, \bibinfo {author}
  {\bibfnamefont {A.~P.}\ \bibnamefont {Mosk}}, \bibinfo {author}
  {\bibfnamefont {J.-M.}\ \bibnamefont {G\'{e}rard}}, \ and\ \bibinfo {author}
  {\bibfnamefont {W.~L.}\ \bibnamefont {Vos}},\ }\href@noop {} {\bibfield
  {journal} {\bibinfo  {journal} {Opt. Lett.}\ }\textbf {\bibinfo {volume}
  {38}},\ \bibinfo {pages} {374} (\bibinfo {year} {2013})}\BibitemShut
  {NoStop}%
\bibitem [{\citenamefont {Notomi}\ \emph {et~al.}(2005)\citenamefont {Notomi},
  \citenamefont {Shinya}, \citenamefont {Mitsugi}, \citenamefont {Kira},
  \citenamefont {Kuramochi},\ and\ \citenamefont {Tanabe}}]{notomitherm}%
  \BibitemOpen
  \bibfield  {author} {\bibinfo {author} {\bibfnamefont {M.}~\bibnamefont
  {Notomi}}, \bibinfo {author} {\bibfnamefont {A.}~\bibnamefont {Shinya}},
  \bibinfo {author} {\bibfnamefont {S.}~\bibnamefont {Mitsugi}}, \bibinfo
  {author} {\bibfnamefont {G.}~\bibnamefont {Kira}}, \bibinfo {author}
  {\bibfnamefont {E.}~\bibnamefont {Kuramochi}}, \ and\ \bibinfo {author}
  {\bibfnamefont {T.}~\bibnamefont {Tanabe}},\ }\href@noop {} {\bibfield
  {journal} {\bibinfo  {journal} {Opt. Express}\ }\textbf {\bibinfo {volume}
  {13}},\ \bibinfo {pages} {2678} (\bibinfo {year} {2005})}\BibitemShut
  {NoStop}%
\bibitem [{\citenamefont {Pan}\ \emph {et~al.}(2008)\citenamefont {Pan},
  \citenamefont {Huo}, \citenamefont {Yamanaka}, \citenamefont {Sandhu},
  \citenamefont {Timp}, \citenamefont {Povinelli}, \citenamefont {Fan},
  \citenamefont {Fejer},\ and\ \citenamefont {Harris}}]{stanfordtherm}%
  \BibitemOpen
  \bibfield  {author} {\bibinfo {author} {\bibfnamefont {J.}~\bibnamefont
  {Pan}}, \bibinfo {author} {\bibfnamefont {Y.}~\bibnamefont {Huo}}, \bibinfo
  {author} {\bibfnamefont {K.}~\bibnamefont {Yamanaka}}, \bibinfo {author}
  {\bibfnamefont {S.}~\bibnamefont {Sandhu}}, \bibinfo {author} {\bibfnamefont
  {L.~S.~R.}\ \bibnamefont {Timp}}, \bibinfo {author} {\bibfnamefont {M.~L.}\
  \bibnamefont {Povinelli}}, \bibinfo {author} {\bibfnamefont {S.}~\bibnamefont
  {Fan}}, \bibinfo {author} {\bibfnamefont {M.~M.}\ \bibnamefont {Fejer}}, \
  and\ \bibinfo {author} {\bibfnamefont {J.~S.}\ \bibnamefont {Harris}},\
  }\href@noop {} {\bibfield  {journal} {\bibinfo  {journal} {Appl. Phys.
  Lett.}\ }\textbf {\bibinfo {volume} {92}},\ \bibinfo {pages} {103114}
  (\bibinfo {year} {2008})}\BibitemShut {NoStop}%
\bibitem [{\citenamefont {Chen}\ \emph {et~al.}(2011)\citenamefont {Chen},
  \citenamefont {Zheng}, \citenamefont {Gu}, \citenamefont {McMillan},
  \citenamefont {Yu}, \citenamefont {Lo}, \citenamefont {Kwong},\ and\
  \citenamefont {Wong}}]{oxydation1}%
  \BibitemOpen
  \bibfield  {author} {\bibinfo {author} {\bibfnamefont {C.~J.}\ \bibnamefont
  {Chen}}, \bibinfo {author} {\bibfnamefont {J.}~\bibnamefont {Zheng}},
  \bibinfo {author} {\bibfnamefont {T.}~\bibnamefont {Gu}}, \bibinfo {author}
  {\bibfnamefont {J.~F.}\ \bibnamefont {McMillan}}, \bibinfo {author}
  {\bibfnamefont {M.}~\bibnamefont {Yu}}, \bibinfo {author} {\bibfnamefont
  {G.}~\bibnamefont {Lo}}, \bibinfo {author} {\bibfnamefont {D.}~\bibnamefont
  {Kwong}}, \ and\ \bibinfo {author} {\bibfnamefont {C.~W.}\ \bibnamefont
  {Wong}},\ }\href@noop {} {\bibfield  {journal} {\bibinfo  {journal} {Opt.
  Expr.}\ }\textbf {\bibinfo {volume} {19}},\ \bibinfo {pages} {12480}
  (\bibinfo {year} {2011})}\BibitemShut {NoStop}%
\bibitem [{\citenamefont {Lee}\ \emph {et~al.}(2009)\citenamefont {Lee},
  \citenamefont {Kiravittaya}, \citenamefont {Kumar}, \citenamefont {Plumhof},
  \citenamefont {Balet}, \citenamefont {Li}, \citenamefont {Francardi},
  \citenamefont {Gerardino}, \citenamefont {Fiore}, \citenamefont {Rastelli},\
  and\ \citenamefont {Schmidt}}]{oxydation2}%
  \BibitemOpen
  \bibfield  {author} {\bibinfo {author} {\bibfnamefont {H.~S.}\ \bibnamefont
  {Lee}}, \bibinfo {author} {\bibfnamefont {S.}~\bibnamefont {Kiravittaya}},
  \bibinfo {author} {\bibfnamefont {S.}~\bibnamefont {Kumar}}, \bibinfo
  {author} {\bibfnamefont {J.~D.}\ \bibnamefont {Plumhof}}, \bibinfo {author}
  {\bibfnamefont {L.}~\bibnamefont {Balet}}, \bibinfo {author} {\bibfnamefont
  {L.~H.}\ \bibnamefont {Li}}, \bibinfo {author} {\bibfnamefont
  {M.}~\bibnamefont {Francardi}}, \bibinfo {author} {\bibfnamefont
  {A.}~\bibnamefont {Gerardino}}, \bibinfo {author} {\bibfnamefont
  {A.}~\bibnamefont {Fiore}}, \bibinfo {author} {\bibfnamefont
  {A.}~\bibnamefont {Rastelli}}, \ and\ \bibinfo {author} {\bibfnamefont
  {O.~G.}\ \bibnamefont {Schmidt}},\ }\href@noop {} {\bibfield  {journal}
  {\bibinfo  {journal} {Appl. Phys. Lett.}\ }\textbf {\bibinfo {volume} {95}},\
  \bibinfo {pages} {191109} (\bibinfo {year} {2009})}\BibitemShut {NoStop}%
\bibitem [{\citenamefont {Riboli}\ \emph {et~al.}(2014)\citenamefont {Riboli},
  \citenamefont {Caselli}, \citenamefont {Vignolini}, \citenamefont {Intonti},
  \citenamefont {Vynck}, \citenamefont {Barthelemy}, \citenamefont {Gerardino},
  \citenamefont {Balet}, \citenamefont {Li}, \citenamefont {Fiore},
  \citenamefont {Gurioli},\ and\ \citenamefont {Wiersma}}]{oxydation3}%
  \BibitemOpen
  \bibfield  {author} {\bibinfo {author} {\bibfnamefont {F.}~\bibnamefont
  {Riboli}}, \bibinfo {author} {\bibfnamefont {N.}~\bibnamefont {Caselli}},
  \bibinfo {author} {\bibfnamefont {S.}~\bibnamefont {Vignolini}}, \bibinfo
  {author} {\bibfnamefont {F.}~\bibnamefont {Intonti}}, \bibinfo {author}
  {\bibfnamefont {K.}~\bibnamefont {Vynck}}, \bibinfo {author} {\bibfnamefont
  {P.}~\bibnamefont {Barthelemy}}, \bibinfo {author} {\bibfnamefont
  {A.}~\bibnamefont {Gerardino}}, \bibinfo {author} {\bibfnamefont
  {L.}~\bibnamefont {Balet}}, \bibinfo {author} {\bibfnamefont
  {L.}~\bibnamefont {Li}}, \bibinfo {author} {\bibfnamefont {A.}~\bibnamefont
  {Fiore}}, \bibinfo {author} {\bibfnamefont {M.}~\bibnamefont {Gurioli}}, \
  and\ \bibinfo {author} {\bibfnamefont {D.}~\bibnamefont {Wiersma}},\
  }\href@noop {} {\bibfield  {journal} {\bibinfo  {journal} {Nat. Mater.}\
  }\textbf {\bibinfo {volume} {13}},\ \bibinfo {pages} {720} (\bibinfo {year}
  {2014})}\BibitemShut {NoStop}%
\bibitem [{\citenamefont {Hennessy}\ \emph {et~al.}(2005)\citenamefont
  {Hennessy}, \citenamefont {Badolato}, \citenamefont {Tamboli}, \citenamefont
  {Petroff}, \citenamefont {Hu}, \citenamefont {Atature}, \citenamefont
  {Dreiser},\ and\ \citenamefont {Imamoglu}}]{chem}%
  \BibitemOpen
  \bibfield  {author} {\bibinfo {author} {\bibfnamefont {K.}~\bibnamefont
  {Hennessy}}, \bibinfo {author} {\bibfnamefont {A.}~\bibnamefont {Badolato}},
  \bibinfo {author} {\bibfnamefont {A.}~\bibnamefont {Tamboli}}, \bibinfo
  {author} {\bibfnamefont {P.~M.}\ \bibnamefont {Petroff}}, \bibinfo {author}
  {\bibfnamefont {E.}~\bibnamefont {Hu}}, \bibinfo {author} {\bibfnamefont
  {M.}~\bibnamefont {Atature}}, \bibinfo {author} {\bibfnamefont
  {J.}~\bibnamefont {Dreiser}}, \ and\ \bibinfo {author} {\bibfnamefont
  {A.}~\bibnamefont {Imamoglu}},\ }\href@noop {} {\bibfield  {journal}
  {\bibinfo  {journal} {Appl. Phys. Lett.}\ }\textbf {\bibinfo {volume} {87}},\
  \bibinfo {pages} {021108} (\bibinfo {year} {2005})}\BibitemShut {NoStop}%
\bibitem [{\citenamefont {Adachi}(2007)}]{adachi}%
  \BibitemOpen
  \bibfield  {author} {\bibinfo {author} {\bibfnamefont {S.}~\bibnamefont
  {Adachi}},\ }\href@noop {} {\bibfield  {journal} {\bibinfo  {journal} {J.
  Appl. Phys.}\ }\textbf {\bibinfo {volume} {102}},\ \bibinfo {pages} {063502}
  (\bibinfo {year} {2007})}\BibitemShut {NoStop}%
\bibitem [{\citenamefont {Levinshtein}, \citenamefont {Rumyantsev},\ and\
  \citenamefont {Shur}(1996)}]{inp}%
  \BibitemOpen
  \bibfield  {author} {\bibinfo {author} {\bibfnamefont {M.}~\bibnamefont
  {Levinshtein}}, \bibinfo {author} {\bibfnamefont {S.}~\bibnamefont
  {Rumyantsev}}, \ and\ \bibinfo {author} {\bibfnamefont {M.}~\bibnamefont
  {Shur}},\ }\href@noop {} {\emph {\bibinfo {title} {Handbook series on
  Semiconductor Parameters}}},\ Vol.~\bibinfo {volume} {1}\ (\bibinfo
  {publisher} {World Scientific},\ \bibinfo {year} {1996})\BibitemShut
  {NoStop}%
\bibitem [{\citenamefont {Anderson}(1989)}]{waterhydrogen}%
  \BibitemOpen
  \bibfield  {author} {\bibinfo {author} {\bibfnamefont {H.}~\bibnamefont
  {Anderson}},\ }\href@noop {} {\emph {\bibinfo {title} {A Physicist's Desk
  Reference}}}\ (\bibinfo  {publisher} {American institute of physics, NY},\
  \bibinfo {year} {1989})\BibitemShut {NoStop}%
\bibitem [{\citenamefont {Tran}\ \emph {et~al.}(2010)\citenamefont {Tran},
  \citenamefont {Combri\'{e}}, \citenamefont {Colman}, \citenamefont {\mbox{De
  Rossi}},\ and\ \citenamefont {Mei}}]{bandfolding}%
  \BibitemOpen
  \bibfield  {author} {\bibinfo {author} {\bibfnamefont {N.}~\bibnamefont
  {Tran}}, \bibinfo {author} {\bibfnamefont {S.}~\bibnamefont {Combri\'{e}}},
  \bibinfo {author} {\bibfnamefont {P.}~\bibnamefont {Colman}}, \bibinfo
  {author} {\bibfnamefont {A.}~\bibnamefont {\mbox{De Rossi}}}, \ and\ \bibinfo
  {author} {\bibfnamefont {T.}~\bibnamefont {Mei}},\ }\href@noop {} {\bibfield
  {journal} {\bibinfo  {journal} {Phys. Rev. B}\ }\textbf {\bibinfo {volume}
  {82}},\ \bibinfo {pages} {075120} (\bibinfo {year} {2010})}\BibitemShut
  {NoStop}%
\bibitem [{\citenamefont {Combri\'{e}}\ \emph {et~al.}(2005)\citenamefont
  {Combri\'{e}}, \citenamefont {Bansropun}, \citenamefont {Lecomte},
  \citenamefont {Parillaud}, \citenamefont {Cassette}, \citenamefont
  {Benisty},\ and\ \citenamefont {Nagle}}]{fabrication}%
  \BibitemOpen
  \bibfield  {author} {\bibinfo {author} {\bibfnamefont {S.}~\bibnamefont
  {Combri\'{e}}}, \bibinfo {author} {\bibfnamefont {S.}~\bibnamefont
  {Bansropun}}, \bibinfo {author} {\bibfnamefont {M.}~\bibnamefont {Lecomte}},
  \bibinfo {author} {\bibfnamefont {O.}~\bibnamefont {Parillaud}}, \bibinfo
  {author} {\bibfnamefont {S.}~\bibnamefont {Cassette}}, \bibinfo {author}
  {\bibfnamefont {H.}~\bibnamefont {Benisty}}, \ and\ \bibinfo {author}
  {\bibfnamefont {J.}~\bibnamefont {Nagle}},\ }\href@noop {} {\bibfield
  {journal} {\bibinfo  {journal} {J. Vac. Sci. Technol. B}\ }\textbf {\bibinfo
  {volume} {23}},\ \bibinfo {pages} {1521} (\bibinfo {year}
  {2005})}\BibitemShut {NoStop}%
\bibitem [{\citenamefont {Tran}\ \emph {et~al.}(2009)\citenamefont {Tran},
  \citenamefont {Combri\'{e}}, \citenamefont {Colman},\ and\ \citenamefont
  {\mbox{De Rossi}}}]{modeconverter}%
  \BibitemOpen
  \bibfield  {author} {\bibinfo {author} {\bibfnamefont {Q.}~\bibnamefont
  {Tran}}, \bibinfo {author} {\bibfnamefont {S.}~\bibnamefont {Combri\'{e}}},
  \bibinfo {author} {\bibfnamefont {P.}~\bibnamefont {Colman}}, \ and\ \bibinfo
  {author} {\bibfnamefont {A.}~\bibnamefont {\mbox{De Rossi}}},\ }\href@noop {}
  {\bibfield  {journal} {\bibinfo  {journal} {Appl. Phys. Lett.}\ }\textbf
  {\bibinfo {volume} {95}},\ \bibinfo {pages} {061105} (\bibinfo {year}
  {2009})}\BibitemShut {NoStop}%
\bibitem [{\citenamefont {Joannopoulos}\ \emph {et~al.}(2008)\citenamefont
  {Joannopoulos}, \citenamefont {Johnson}, \citenamefont {Winn},\ and\
  \citenamefont {Meade}}]{joannopoulos}%
  \BibitemOpen
  \bibfield  {author} {\bibinfo {author} {\bibfnamefont {J.}~\bibnamefont
  {Joannopoulos}}, \bibinfo {author} {\bibfnamefont {S.~G.}\ \bibnamefont
  {Johnson}}, \bibinfo {author} {\bibfnamefont {J.~N.}\ \bibnamefont {Winn}}, \
  and\ \bibinfo {author} {\bibfnamefont {R.~D.}\ \bibnamefont {Meade}},\
  }\href@noop {} {\emph {\bibinfo {title} {Photonic Crystals Molding the flow
  of light}}}\ (\bibinfo  {publisher} {PRINCETON UNIVERSITY PRESS},\ \bibinfo
  {year} {2008})\BibitemShut {NoStop}%
\bibitem [{\citenamefont {Oskooi}\ \emph {et~al.}(2010)\citenamefont {Oskooi},
  \citenamefont {Roundy}, \citenamefont {Ibanescu}, \citenamefont {Bermel},
  \citenamefont {Joannopoulos},\ and\ \citenamefont {Johnson}}]{meep}%
  \BibitemOpen
  \bibfield  {author} {\bibinfo {author} {\bibfnamefont {A.~F.}\ \bibnamefont
  {Oskooi}}, \bibinfo {author} {\bibfnamefont {D.}~\bibnamefont {Roundy}},
  \bibinfo {author} {\bibfnamefont {M.}~\bibnamefont {Ibanescu}}, \bibinfo
  {author} {\bibfnamefont {P.}~\bibnamefont {Bermel}}, \bibinfo {author}
  {\bibfnamefont {J.~D.}\ \bibnamefont {Joannopoulos}}, \ and\ \bibinfo
  {author} {\bibfnamefont {S.~G.}\ \bibnamefont {Johnson}},\ }\href@noop {}
  {\bibfield  {journal} {\bibinfo  {journal} {Comp. Phys. Comm.}\ }\textbf
  {\bibinfo {volume} {181}},\ \bibinfo {pages} {687} (\bibinfo {year}
  {2010})}\BibitemShut {NoStop}%
\bibitem [{Note1()}]{Note1}%
  \BibitemOpen
  \bibinfo {note} {MIT material property database,
  http://www.mit.edu/~6.777/matprops/matprops.htm}\BibitemShut {NoStop}%
\bibitem [{\citenamefont {Schubert}\ \emph {et~al.}(1995)\citenamefont
  {Schubert}, \citenamefont {Gottschalch}, \citenamefont {Herzinger},
  \citenamefont {Yao},\ and\ \citenamefont {Snyder}}]{schubert}%
  \BibitemOpen
  \bibfield  {author} {\bibinfo {author} {\bibfnamefont {M.}~\bibnamefont
  {Schubert}}, \bibinfo {author} {\bibfnamefont {V.}~\bibnamefont
  {Gottschalch}}, \bibinfo {author} {\bibfnamefont {C.~M.}\ \bibnamefont
  {Herzinger}}, \bibinfo {author} {\bibfnamefont {H.}~\bibnamefont {Yao}}, \
  and\ \bibinfo {author} {\bibfnamefont {P.~G.}\ \bibnamefont {Snyder}},\
  }\href@noop {} {\bibfield  {journal} {\bibinfo  {journal} {J. Appl. Phys.}\
  }\textbf {\bibinfo {volume} {77}},\ \bibinfo {pages} {3416} (\bibinfo {year}
  {1995})}\BibitemShut {NoStop}%
\bibitem [{\citenamefont {Palik}(1985)}]{sin}%
  \BibitemOpen
  \bibfield  {author} {\bibinfo {author} {\bibfnamefont {E.}~\bibnamefont
  {Palik}},\ }\href@noop {} {\emph {\bibinfo {title} {Handbook of Optical
  Constants of Solids}}}\ (\bibinfo  {publisher} {Academic Press},\ \bibinfo
  {year} {1985})\BibitemShut {NoStop}%
\end{thebibliography}
\end{document}